# Detection of sub-lattice magnetism in sigma-phase Fe-V compounds by zero-field NMR

S. M. Dubiel[1*], J. R. Tozoni[2], J. Cieślak[1], D. C. Braz[2], E.L. G. Vidoto[2] and T. J. Bonagamba[2]

[1]Faculty of Physics and Computer Science, AGH University of Science and Technology, PL-30-059 Kraków, Poland, and [2]Instituto de Física de São Carlos, Universidade de São Paulo, Caixa Postal 369, São Carlos, 13560-970, São Paulo, Brazil.

The first successful measurements of a sub-lattice magnetism with $^{51}$V NMR techniques in the sigma-phase $Fe_{100-x}V_x$ alloys with x = 34.4, 39.9 and 47.9 are reported. Vanadium atoms present on all five crystallographic sites are magnetic. Their magnetic properties are characteristic of a given site, which strongly depend on the composition. The strongest magnetism exhibit sites A and the weakest one sites D. The estimated average magnetic moment per V atom decreases from 0.36 $\mu_B$ for $x$ = 34.4 to 0.20 $\mu_B$ for $x$ = 47.9. The magnetism revealed at V atoms is linearly correlated with the magnetic moment of Fe atoms, which implies that the former is induced by the latter.

A sigma-phase can be produced by a solid-state reaction in some alloy systems in which at least one constituent is a transition element. It has a tetragonal crystallographic structure and its unit cell contains 30 atoms that are distributed over five different crystallographic sites A, B, C, D and E. Because of the high coordination numbers (12-15), the phase belongs to a family of the so-called Frank-Kasper phases. Among over 50 binary alloys in which the sigma-phase was found, only that in the Fe-Cr and Fe-V alloys has well evidenced magnetic properties [1-6]. Despite first magnetic investigations of the sigma-phase were carried out over 40 years ago [2,3], its magnetism, which is usually termed as weak and low temperature, is not well understood. Some features like a lack of saturation of magnetisation even in an external magnetic field of 15 T [4-6] and the Rhodes-Wohlfarth criterion speak in favor of its itinerant character, but it remains fully unknown (a) if both kinds of the constituting atoms contribute to the magnetism, (b) what are the values of the magnetic moments localized at the constituting atoms occupying different crystallografic sites and (c) is the magnetic structure co-linear or not. The difficulty in answering these questions arrises on one hand from a failure to produce a big enough single-crystal of the sigma-phase that could be used in a neutron-diffraction experiment to decifer the magnetic structure, and on the other, in a lack of high-enough resolution of the Mössbauer Spectroscopy (MS), which partly follows from the weak magnetism of the sigma-phase and partly from its complex crystallographic structure combined with a chemical disorder of atom distribution over the five sites [7].

In this Communication we report the first successful measurement of the nuclear magnetic resonance (NMR) spectra without applying external magnetic field (so called zero-field NMR) [8-11] on the σ-phase $Fe_{100-x}V_x$ samples with x = 34.4, 39.9 and 47.9, which gives a clear evidence that V atoms occupy five sites and have a non-zero spin-density (hyperfine field) that is characteristic of a given site.

The samples of the σ-phase used in this study were prepared by an isothermal annealing of master ingots of the α-phase at 973 K for 25 days. More detail description of the fabrication process and the verification of the final phase and its chemical composition are given elsewhere [6]. For NMR measurements, the samples were in form of powder (~100 mg) obtained by attrition bulk samples in an agate mortar and, afterwards, mixed with paraffin.

The zero-field NMR experiments were carried out using a Discovery Tecmag Console, which operates in the frequency range of 1 to 600 MHz. The spectra were obtained by exciting the nuclei frequency-by-frequency (0.3125-MHz step), within a frequency range of 10 to 100 MHz, and measuring the respective modules of the complex echo signals acquired.

For the samples $Fe_{65.6}V_{34.4}$ and $Fe_{60.1}V_{39.9}$, the echoes were obtained after the application of two 1-μs radiofrequency pulses, separated by a fixed delay of 20 μs. The repetition time for the spin-echo experiments was set to 50 ms and the number of averages to 500 scans. However, for the sample $Fe_{52.1}V_{47.9}$, due to the low signal-to-noise ratio, the echoes were obtained after the application of two 2-μs radiofrequency pulses, the inter-pulse delay was set to 7 μs, the repetition time to 10 ms and the number of averages to 150000 scans. All the measurements were carried out at 4.2 K.

The NMR signal can in this case originate from both $^{51}V$ and $^{57}Fe$ nuclei. However, the main lines observed in the spectra (Figures 1 and 2) were assigned to the $^{51}V$ nuclei due to the following three main reasons: i) much higher natural abundance of the $^{51}V$ nuclei (99.8%) as compared to that of $^{57}Fe$ (2.2%); ii) more intense giromagnetic factor of $^{51}V$ nuclei (11.2 MHz/T) in contrast with that of $^{57}Fe$ (1.4 MHz/T); and iii) Mössbauer experiments carried out for similar samples to those studied in this work indicate that local hyperfine fields are dispersed from 5 to 20 T at 4.2 K, with an average value of about 13 T [6]. Taking into consideration the two last reasons, one would expect to observe $^{57}Fe$ lines in the NMR spectrum in the frequency range of 5 to 30 MHz, while the recorded spectra are observed from 10 to 100 MHz. Consequently, one could conclude that all the observed lines are associated with $^{51}$V nuclei. Additionally, preliminary NMR measurements performed at the center of each peak (A through E) of the spectrum observed for sample $Fe_{65.6}V_{34.4}$ indicated that they present quadrupolar oscillations, a specific behavior of quadupolar nuclei. In our case, only $^{51}$V nuclei are quadrupolar [9].

Due to the fact that the NMR spectrometer was used for detecting very wide spectra, special care was taken to avoid spurious artifacts affecting the wide line spectral shape from transmitter, receiver, cables, radiofrequency (rf) probe, etc. In order to circumvent all the undesirable contributions, the radiofrequency probe coil was designed to avoid self-resonances in the experiment frequency range (10 to 100 MHz) and its connections to the spectrometer were made without using tuning/matching by capacitors and transmitter/receiver duplexing with quarter-wavelength-cable. Before the experiments, the following calibration procedures were carried out: i) the rf probe was fully characterized by the use of a vector impedance meter; ii) the spectrometer frequency response, without probe, was analyzed by connecting the transmitter directly to the receiver and running the experiment over the full experiment frequency range; and iii) the same procedure used in ii) was repeated with the inclusion of the rf probe. After these procedures, it was possible to observe all the spurious contributions of the spectrometer to the spectra, allowing to safely separate/eliminate them from the observed spectra. Figure 1 shows the raw spectrum observed for the sample $Fe_{65.6}V_{34.4}$ together with the most representative calibration curve, where one can easily observe the spurious artifacts introduced by all the components of the spectrometer. After quickly identifying the spurious peaks, they were simply removed from the spectrum by extracting/moving the corresponding points from the spectrum data (see asterisks), without any additional data manipulation. In the cases were the signal-to-noise was not good enough to distinguish the signal from the noise for both calibration curve and NMR data, the spectrum intensity was kept the same.

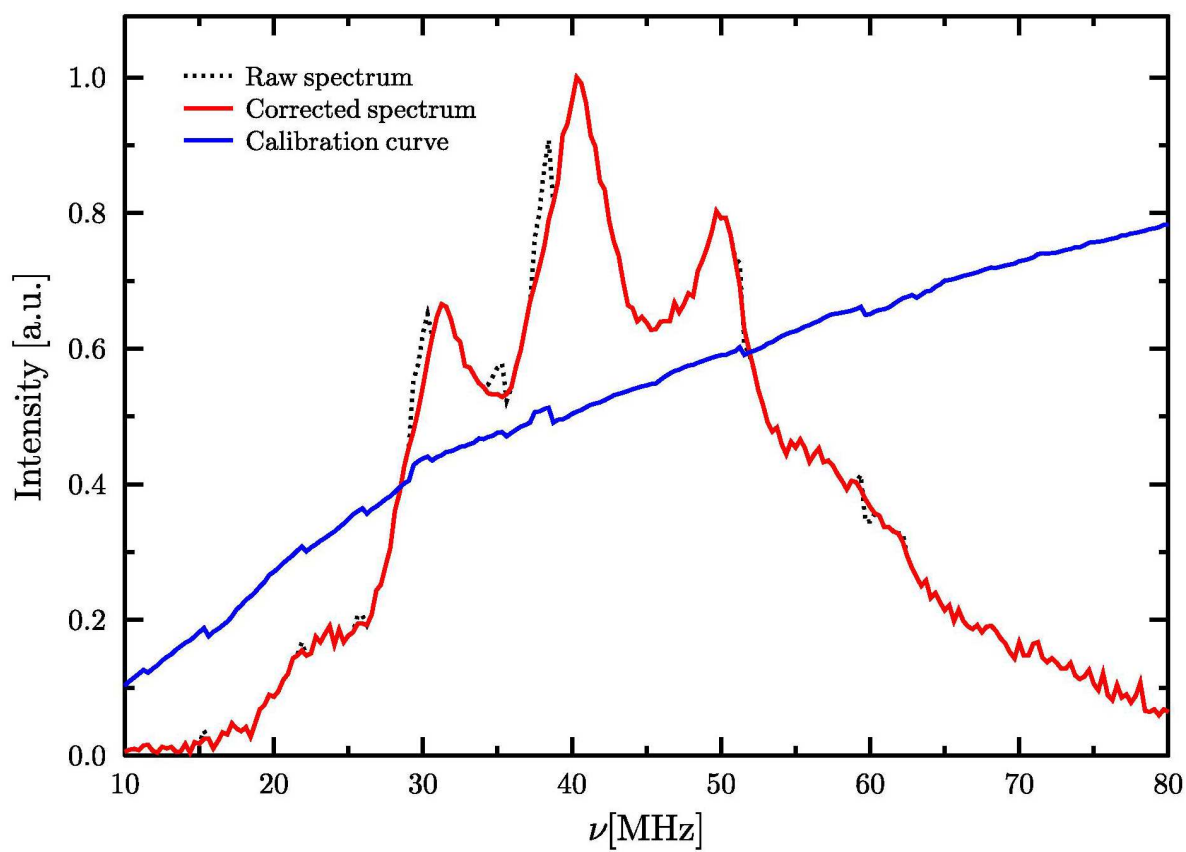

Fig. 1 (Color online) Raw $^{51}$V NMR spectrum observed for the sample $Fe_{65.6}V_{34.4}$ together with a calibration curve and corrected spectrum.

In order to get the final shape for the spectra, the intensities were corrected to compensate the frequency dependent sensitivity of the NMR system, which was also experimentally estimated. Fig. 2 shows the final spectra for all three samples, taking the same procedures discussed above to insure that the spectral features are predominately due to the $^{51}$V hyperfine interactions.

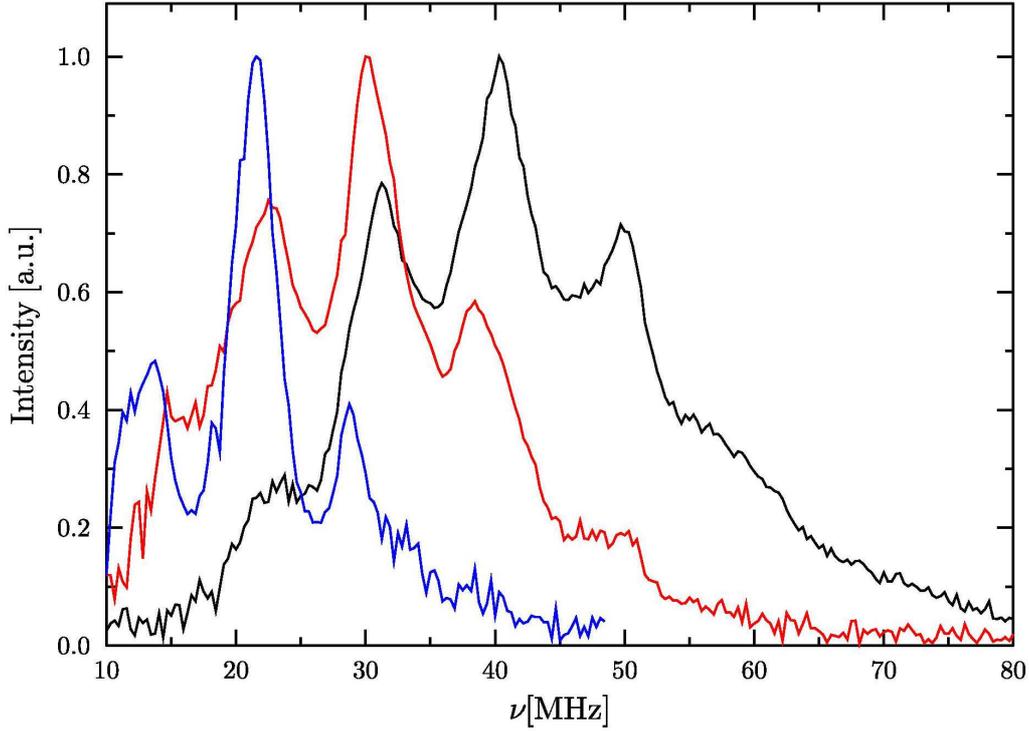

Fig. 2 (Color online) $^{51}$V NMR corrected spectra recorded at 4.2 K for Fe-content *(1-x)* = 65.6, *(1-x)* = 60 and *(1-x)* = 52.1, in the sequence from right to left.

With the purpose of estimating the area of each line of the spectra, they were decomposed by a multiple-Lorentzian fitting procedure, using 5, 5 and 3 peaks for samples $Fe_{65.6}V_{34.4}$, $Fe_{60.1}V_{39.9}$ and $Fe_{52.1}V_{47.9}$, respectively (see Fig. 3), assuming they should consist basically of five resonance lines corresponding to the different crystallographic sites A, B, C, D and E, superimposed on a background. Despite trying to keep free all the fitting parameters during the spectral decomposition procedure, the line widths were kept as similar as possible and the correlation among these areas and the respective site populations got from neutron diffraction [7] was kept in mind. Due to this reason, errors of about 10% for the line intensities (areas), which include, not only the error indicated by the fittings but also the line width variability should be noticed. The shoulder observed in the higher frequency side of the $Fe_{65.6}V_{34.4}$ sample's spectrum (~70 MHz), which was associated to spurious signals, and neglected in the fitting procedures, disappears for the other samples or eventually hides under the line A. In this case, mostly for the sample $Fe_{60.1}V_{39.9}$, this could give an additional error to the line A intensity, since only 5 peaks were considered in the fitting procedure.

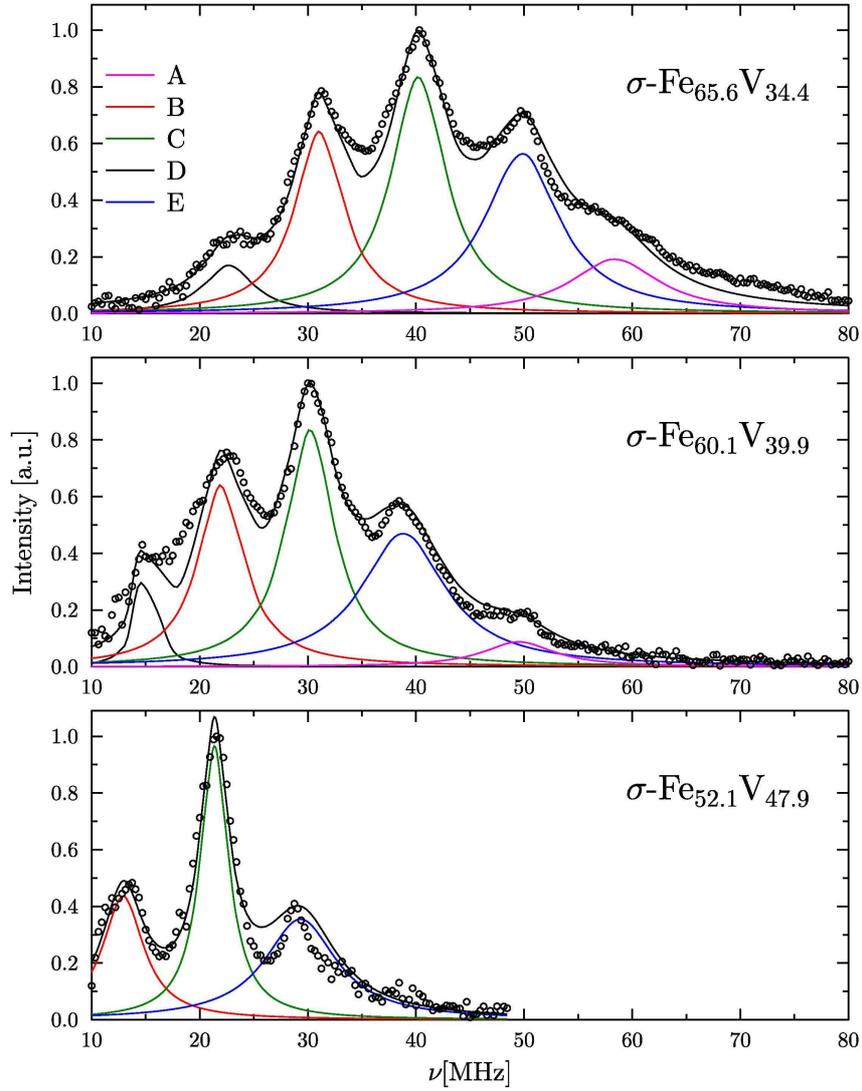

Fig. 3 (Color online) Corrected (circles) and deconvoluted spectra (full lines) recorded at 4.2 K for the investigated samples.

The best-fit parameters obtained with the above-described procedure are given in Table I. Taking in consideration the line intensities and the respective site populations got from the neutron diffraction [7], one can obtain a good correlation between the two experimental methods, as indicated by the correlation coefficient $R^2$ in Table II.

Additional correlations were found for the NMR line positions and the magnetic moment per Fe atom versus the Fe content as shown in Figures 2, 4 and 5, confirming that the bigger the iron concentration the higher the local magnetic field in the V sites.

Before a more detailed discussion of the results will proceed, let us first notice that a well-defined five-line structure seen in the intensity of the measured spectra is the characteristic feature. This means that (1) the hyperfine field (spin-density) exists on V atoms occupying all five crystallographic sites, and (2) it is characteristic of a given site. This agrees, at least, qualitatively with the observation and calculation known for disordered α-Fe-V alloys, where V atoms in Fe-rich alloys have magnetic moments of the order of $1\mu_B$ [12-14].

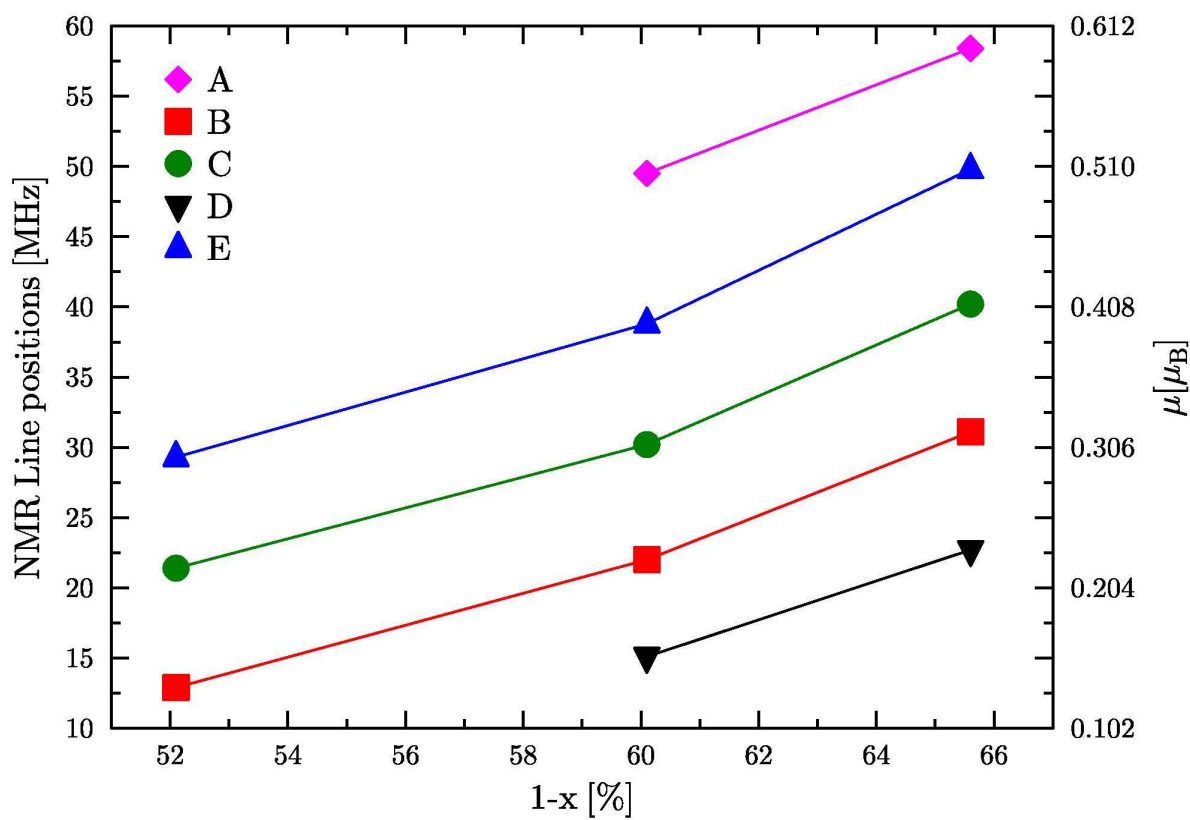

Fig. 4 (Color online) Resonance line positions for particular lattice sites versus Fe content, *(1-x)*. The data are connected to guide the eye. The right-hand-side y-axis has been scaled in the corresponding magnetic moment estimated using the scaling factor as given elsewhere [12,13].

Let us first discuss the question of the effect of composition on the peak position of each of the five resonance lines. As shown in Fig. 3, all three spectra are shifted with *(1-x)* towards higher frequencies (hyperfine fields). Concerning positions of particular resonance lines, a strong quasi-linear increase with Fe content, *(1-x)*, can be observed for all five sites – see Fig. 4. The increase, which is slightly enhanced for higher Fe-concentrations, is rather site independent. The average shift between the subsequent resonance lines has the following approximate values: 8.93 MHz or 0.80 T for *x* = 34.4; 8.60 MHz or 0.77 T for x = 39.9 and, finally, 8.20 MHz or 0.73 T for *x* = 47.9. These figures can be rescaled into the underlying magnetic moments. For that purpose, the scaling constant of 9 T/$\mu_B$ deduced from the data published elsewhere [12,13] can be used. By doing so, the following shifts expressed in Bohr magneton have been obtained: 0.0885 for *x* = 34.4, 0.0883 for *x* = 39.9 and 0.0813 for *x* = 47.9, i.e. they hardly depend on composition. In order to compare this behavior with that of the average magnetic moment per Fe atom, $<\mu>$, as determined elsewhere [6], a weighted average frequency (gravity center), $<\nu>$, of each spectrum was calculated and the relationship between them and Fe-contents is displayed in Fig. 5. As can be here seen, both $<\nu>$ and $<\mu>$ are linearly dependent on *(1-x)*, indicating that the hyperfine field (spin-density) revealed at $^{51}$V nuclei has been induced by magnetic Fe atoms. This observation is consistent with the itinerant character of magnetism of the σ-FeV alloys.

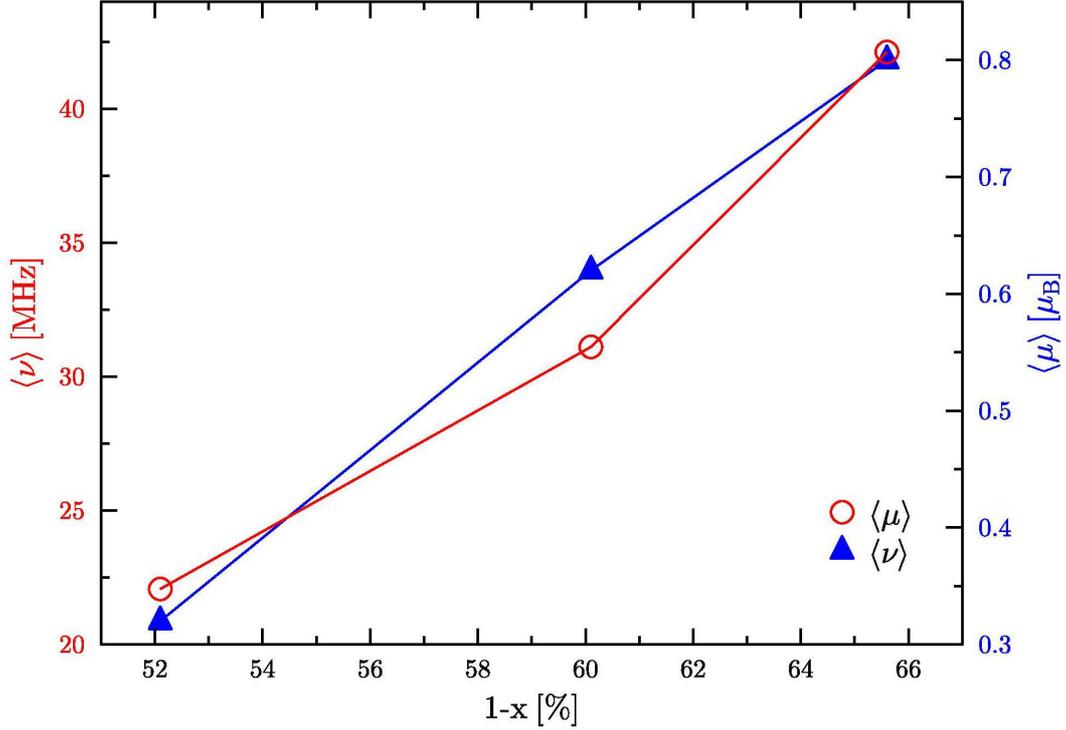

Fig. 5 (Color online) Relationship between the average resonance frequency, $<\nu>$, of the measured NMR spectra, the average magnetic moment per Fe atom, $<\mu>$, and Fe content, $(1-x)$. The lines connecting the data are drawn to guide the eye.

In summary, we have succeeded to record for the first time NMR spectra on magnetic σ-phase samples. The spectra give a clear evidence that in σ-FeV alloys V atoms occupy all five sublattices and all of them have non-zero hyperfine field, hence a magnetic moment whose value strongly depends on the site, and for a given site on sample composition. The strongest magnetism show A sites with the hyperfine field (magnetic moment) of 5.21 T (0.58 $\mu_B$) for $x = 34.4$ and 4.42 T (0.49 $\mu_B$) for $x = 39.9$, and the weakest one sites D having the hyperfine field (magnetic moment) of 2.03 T (0.225 $\mu_B$) for $x = 34.4$ and 1.35 T (0.15 $\mu_B$) for $x = 39.9$. For the highest V-content the resonance lines due to these two sites have been not detected. The average hyperfine field (magnetic moment) at $^{51}$V nucleus decreases from $<B> = 3.76$ (0.42 $\mu_B$) for $x = 34.4$ to 1.97 T (0.22 $\mu_B$) for $x = 47.9$. It is clear from these data and the linear correlation between $<\nu>$ and $<\mu>$, that the magnetism observed on V atoms strongly depends on the composition and it is induced by that of Fe atoms.

Additional $^{57}$Fe-enriched samples and experiments are now being implemented in order to measure signals for $^{57}$Fe nuclei and quadrupolar oscillations [9] for all the five $^{51}$V sites, which will give additional important structural information about these materials from the NMR point of view.

* Corresponding author: dubiel@novell.ftj.agh.edu.pl (S. M. Dubiel)

**Acknowledgements**

Authors acknowledge Prof. Jair C. C. de Freitas for the discussions about this study. The project was partly supported by the Ministry of Science and Higher Education, Warsaw and the Brazilian Science Foundations FAPESP, CAPES and CNPq.

**Table I**

Best-fit parameters as obtained for $^{51}$V spectra recorded at 4.2 K on the investigated samples.

| Site | Line position [MHz] | | | Line separation [MHz] | | |
|---|---|---|---|---|---|---|
| | x = 34.4 | x = 39.9 | x = 47.9 | x = 34.4 | x = 39.9 | x = 47.9 |
| D | 22.70 | 15.10 | - | - | - | - |
| B | 31.10 | 22.00 | 12.91 | 8.40 | 6.90 | - |
| C | 40.25 | 30.20 | 21.36 | 9.15 | 8.20 | 8.45 |
| E | 49.80 | 38.83 | 29.32 | 9.55 | 8.63 | 7.96 |
| A | 58.40 | 49.50 | - | 8.60 | 10.67 | - |
| Average line separation [MHz] | | | | 8.93 | 8.60 | 8.20 |
| Average frequency [MHz] | | | | 42.12 | 31.11 | 22.07 |

**Table II**

Relative probabilities, *P*, of finding V atoms at different sites A, B, C, D and E in a unit cell of the σ-phase samples of $Fe_{100-x}V_x$ compounds as derived from the measured $^{51}V$ NMR spectra and those, in brackets, obtained from neutron diffractions studies [7]. The linear correlation coefficient between the two series of *P*. is denoted as $R^2$.

| Site | x = 34.4 | x = 39.9 | x = 47.9 |
|---|---|---|---|
|  | P [%] | P [%] | P [%] |
| A | 5.8 (0.6) | 5.5 (0.8) | - (1.0) |
| B | 21.8 (26.0) | 24.3 (24.7) | 25.6 (23.3) |
| C | 31.5 (35.1) | 33.0 (36.3) | 38.4 (36.8) |
| E | 29.2 (36.5) | 32.1 (36.3) | 36.0 (36.8) |
| D | 11.7 (1.8) | 5.1 (1.9) | - (2.1) |
| $R^2$ | 0.86 | 0.96 | 0.95 |